\documentclass[a4paper,11pt]{article}
\usepackage{pos}

\usepackage[capitalize]{cleveref}
\usepackage{wrapfig}

\title{Probing hadronic interactions with high-energy and low-energy muons in extensive air showers}
\ShortTitle{Probing hadronic interactions with high-energy and low-energy muons in EASs}

\author*[a]{Lilly Pyras}
\author[a]{Dennis Soldin}
\author[b]{Felix Riehn}

\affiliation[a]{Department of Physics and Astronomy, University of Utah, UT 84112, USA}
\affiliation[b]{Department of Physics, TU Dortmund University, 44221 Dortmund, Germany}

\emailAdd{lilly.pyras@utah.edu}
\emailAdd{dennis.soldin@utah.edu}
\emailAdd{felix.riehn@tu-dortmund.de}

\abstract{
High-energy cosmic rays are observed indirectly by detecting the extensive air showers initiated in Earth’s atmosphere. The interpretation of experimental data relies on accurate modeling of the air shower development. Simulations based on current hadronic interaction models show significant discrepancies with measurements of the muon content in air showers, commonly referred to as the Muon Puzzle, indicating severe shortcomings in the understanding of particle physics. A hybrid detector design with a surface array and deep underground detector, such as the IceCube Neutrino Observatory, allows simultaneous measurements of muons at two vastly different energies: at GeV energies and above a few 100\,GeV. We present phenomenological studies of low-energy and high-energy muons in simulated air showers, and discuss how hybrid measurements can provide constraints on multi-particle production in hadronic interaction models.
}

\ConferenceLogo{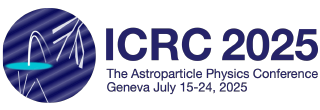}

\FullConference{39th International Cosmic Ray Conference (ICRC2025)\\
 15–24 July 2025\\
Geneva, Switzerland\\}

\begin{document}
\maketitle

\section{Introduction} 

Cosmic rays are atomic nuclei that penetrate the Earth’s atmosphere with energies exceeding 100\,EeV --- corresponding to center-of-mass energies more than ten times higher than those reached at the LHC. When cosmic rays interact in the atmosphere, they produce a cascade of secondary particles, known as extensive air showers (EASs), which can be measured by ground-based detector arrays. The interpretation of these experimental data relies on accurate modeling of the air shower development. Despite high-precision measurements of the cosmic-ray energy spectrum across many orders of magnitude, their sources remain unknown, the mechanisms behind their acceleration are still uncertain, their mass composition is not well constrained, and several features observed in the energy spectrum are not well understood~\cite{Coleman:2022abf}. 

Many uncertainties persist due to discrepancies between the observed number of muons in extensive air showers and predictions from simulations --- a phenomenon commonly referred to as the \emph{Muon Puzzle}. In particular, analyses of data from the Pierre~Auger~Observatory have shown an excess in the number of muons compared to simulations~\cite{PierreAuger:2014ucz, PierreAuger:2016nfk,PierreAuger:2024neu}. A systematic meta-analysis of data from nine different air-shower experiments has revealed an energy-dependent trend in these discrepancies, with high statistical significance~\cite{Dembinski:2019uta,Soldin:2021wyv,Cazon:2020zhx}. Recent studies suggest that these discrepancies may indicate severe deficits in our understanding of particle physics~\cite{Albrecht:2021cxw,ArteagaVelazquez:2023fda}.

The main challenge in modeling extensive air showers and their muon content is the treatment of hadronic interactions in the atmosphere across many orders of magnitude in energy. Since EAS development is largely governed by soft-QCD processes, which cannot be derived from first principles, current hadronic interaction models rely heavily on extrapolations into poorly constrained regions of phase space. Although hadronic interactions have been studied at collider experiments up to several TeV, those at higher energies remain inaccessible to current facilities. Additionally, interactions in the forward region and the specific combinations of particles involved --- both of which play a critical role in particle production in EASs --- also lie beyond the reach of existing experimental capabilities. Current measurements of high-energy extensive air showers typically probe low-energy muons with energies of a few GeV, which are mainly produced in low-energy interactions during the shower development. However, the muon spectrum at higher energies is not well measured. Additionally, due to limited experimental data on EASs, it is uncertain at which shower energies the discrepancies between simulations and measurements arise.
Further measurements from both collider and EAS experiments are essential to advance our understanding of particle production in hadronic interactions~\cite{Albrecht:2021cxw}.

A unique opportunity to study muon production in hadronic interactions is the simultaneous measurement of both GeV and TeV muons in extensive air showers, which is possible with hybrid lerge-scale detectors, such as the IceCube Neutrino Observatory (IceCube)~\cite{Aartsen:2016nxy}. While IceCube has demonstrated its ability to measure the multiplicity of high-energy muons ($E_\mu>500$\,GeV) on an event-by-event basis~\cite{Verpoest:2023qmq}, the most recent measurement of low-energy muons ($E_\mu\simeq1$\,GeV) with IceTop~\cite{IceCubeCollaboration:2022tla} --- the surface array of IceCube --- is based on a statistical analysis which does not provide information on event-by-event basis. 
Nonetheless, promising new reconstruction techniques are currently being developed that will enable simultaneous event-by-event measurements of the high- and low-energy muon content in EASs in the near future~\cite{IceCube:2023suf,Mark:ICRC2025,Lincoln:ICRC2025, LillyICRC25}.

\newpage
The analysis of muons at two vastly different energies in the same air shower provides important information about the energy sharing between low-energy and high-energy interactions during the EAS development. This approach also allows investigation into whether the Muon Puzzle extends to muons in the TeV region. Such measurements can thereby determine if the observed discrepancies originate from unconventional processes in the forward region within the Standard Model, or if fundamentally new processes are required to describe the observed muons in EAS.

\begin{figure}[!t]
\vspace{-1.em}
\mbox{\hspace{-1.6em}\includegraphics[width=0.525\textwidth]{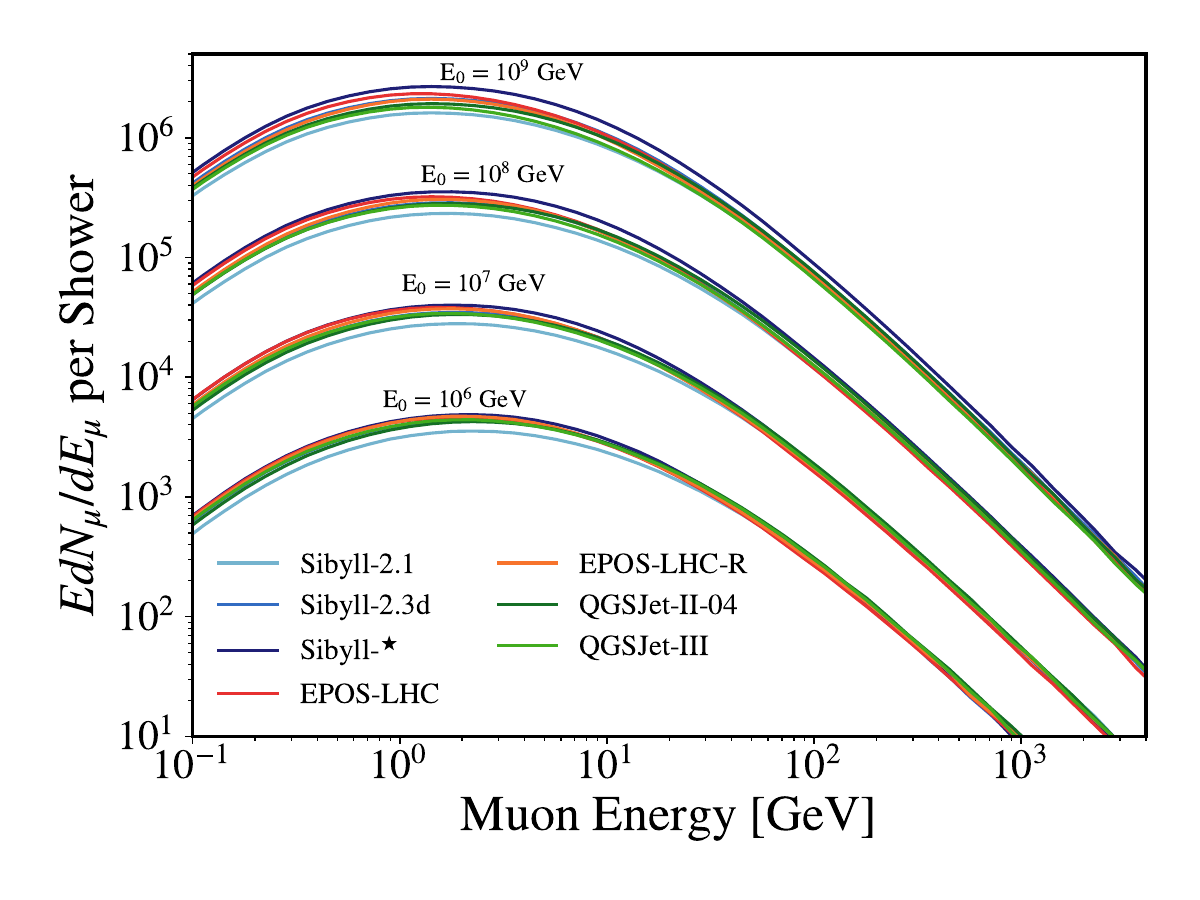}
\includegraphics[width=0.525\textwidth]{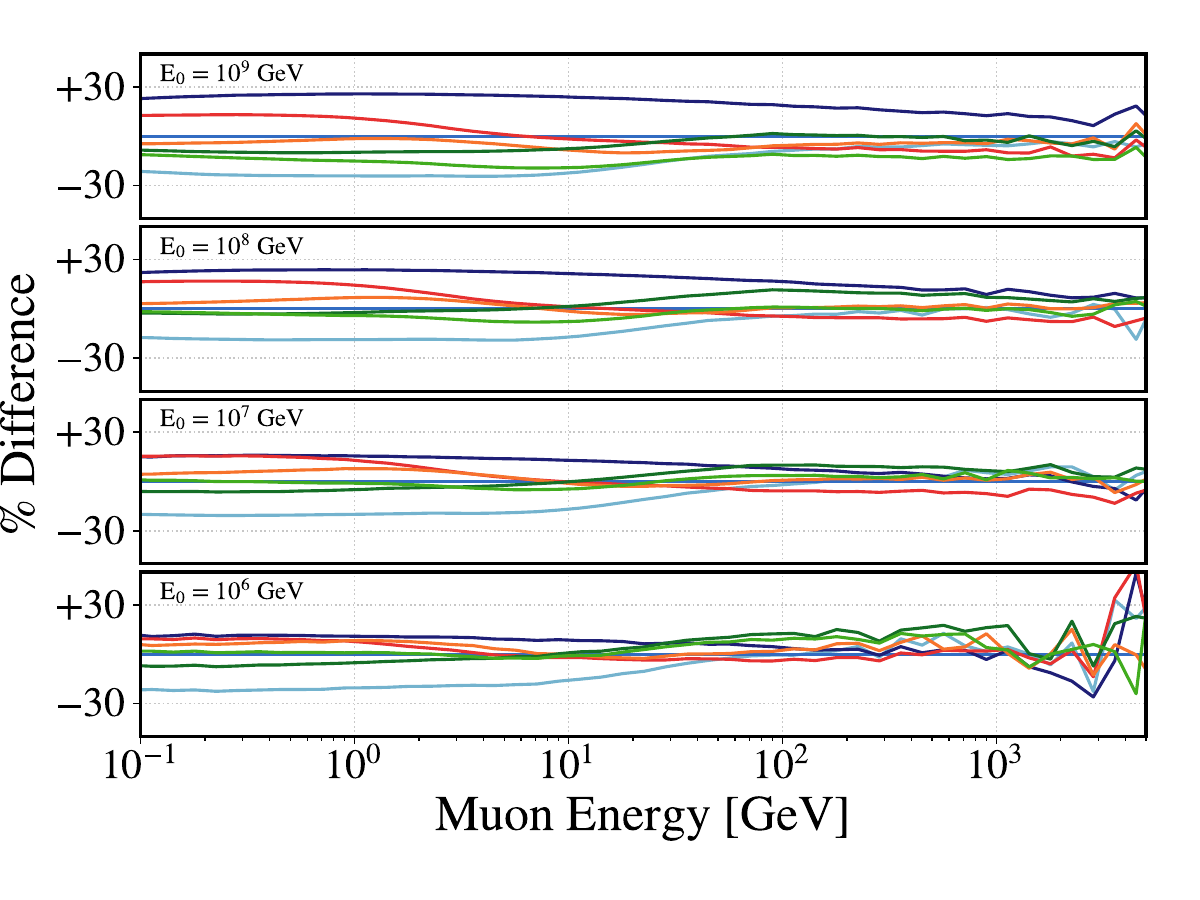}}
\vspace{-2.em}
\caption{Average muon energy spectra (left) obtained from CORSIKA simulations for hydrogen, helium, oxygen, and iron primaries at $10^6$, $10^7$, $10^8$, $10^9$\,GeV and $\cos(\theta)=1$, using various hadronic interaction models as indicated in the legend. The primary mass fractions from the GSF model~\cite{Dembinski:2017zsh} are used. Also shown are the model differences with respect to Sibyll\,2.3d for each primary energy separately (right).}
\label{fig:muon_spectrum}
\end{figure}

We will explore the opportunities for studying multi-particle production in EASs by simultaneously measuring low-energy and high-energy muons. In \cref{sec:Nmu} we will present Monte-Carlo studies of the muon content in EASs based on various hadronic interaction models, including most recent models. We will show how simultaneous measurements of muons in two energy regimes can provide unique information to constrain hadronic EAS models. The opportunity to additionally measure the muon production depth will be explored in \cref{sec:MPD}. We will show how simultaneous measurements of multiple air-shower observables provide unique tests of hadronic models.

\section{Number of Muons in EASs} 
\label{sec:Nmu}
To study the muon content in air showers in detail, EAS simulations for proton, helium, oxygen and iron nuclei were produced using the CORSIKA\,7 software~\cite{Heck:1998vt}, based on multiple hadronic interaction models. For each nucleus, zenith angle ($\cos\theta = 1$ and $\cos\theta = 0.85$), and hadronic model, 100 showers were simulated at primary energies of $10^6$, $10^7$, and $10^8$ GeV, and 10 showers at $10^9$ GeV. The observation level is chosen to be at an elevation of $2840$\,m (around $690$\,g/cm$^2$) and an atmospheric profile for the South Pole in April is used, corresponding to IceCube conditions.
The cosmic-ray mass composition is assumed following the GSF flux model~\cite{Dembinski:2017zsh}, a data-driven approach that also provides flux uncertainties.

\Cref{fig:muon_spectrum} (left) shows the average energy spectrum of muons in vertical air showers. The spectra are shown for the hadronic interaction models Sibyll\,2.1~\cite{Ahn:2009wx}, Sibyll\,2.3d~\cite{Riehn:2019jet}, EPOS-LHC~\cite{Pierog:2013ria}, EPOS-LHC-R~\cite{Pierog:2023ahq}, QGSJET-II-04~\cite{Ostapchenko:2013pia,Ostapchenko:2005nj}, and QGSJet-III~\cite{Ostapchenko:2024myl}. 
Also shown are predictions from the model Sibyll$^\bigstar$~\cite{Riehn:2024prp} which is a series of phenomenologically modified versions of Sibyll\,2.3d which aim to increase muon production in order to provide a possible solution to the Muon Puzzle. This is done by enhancing the production of baryon-antibaryon pairs, kaons, or neutral rho-mesons, or using a mix of baryon-antibaryon and neutral rho-meson enhancements. Throughout this work, the baryon-antibaryon and neutral rho-enhancement mix is used. 
The model predictions show differences in the spectrum, ranging from a few percent at high muon energies, up to around $50$\% for low-energy muons, as demonstrated in \cref{fig:muon_spectrum} (right). The differences at low muon energies increase with increasing EAS energy.

\begin{figure}[!t]
\centering
\vspace{-1em}
\includegraphics[width=\textwidth]{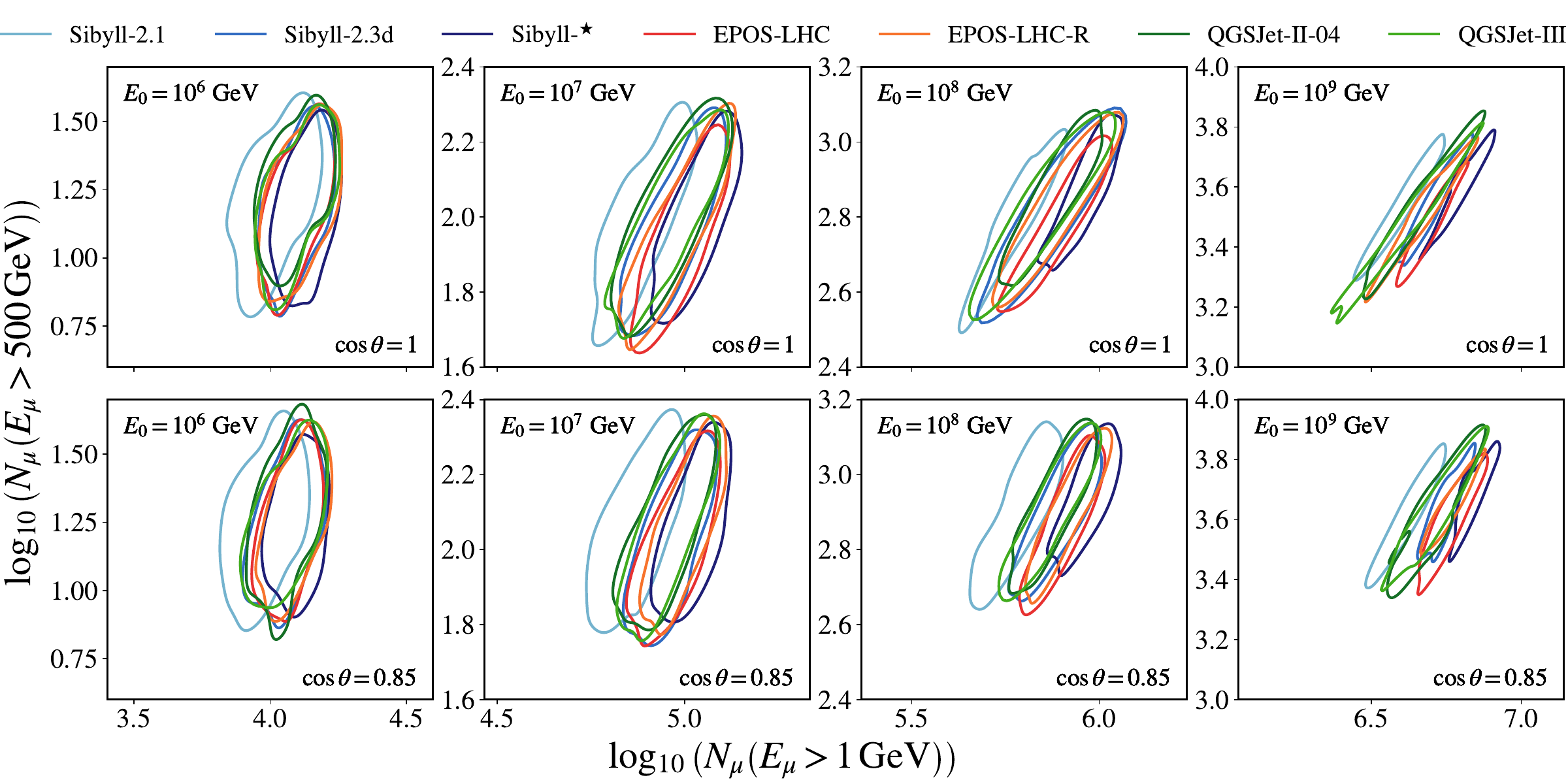}
\caption{68.3\% contours (derived from kernel density estimation) of the number of muons at ground, $N_\mu$, with energies $E_\mu>1$\,GeV as a function of the number of muons at ground with $E_\mu>500$\,GeV, per simulated air shower, obtained from the different hadronic models. The distributions are weighted to GSF and shown for four different EAS energies and $\cos(\theta)=1$ (top) and $\cos(\theta)=0.85$ (bottom).}
\label{fig:contour_GSF}
\end{figure}

\begin{figure}[!t]
\vspace{-1em}
\centering
\includegraphics[width=\textwidth]{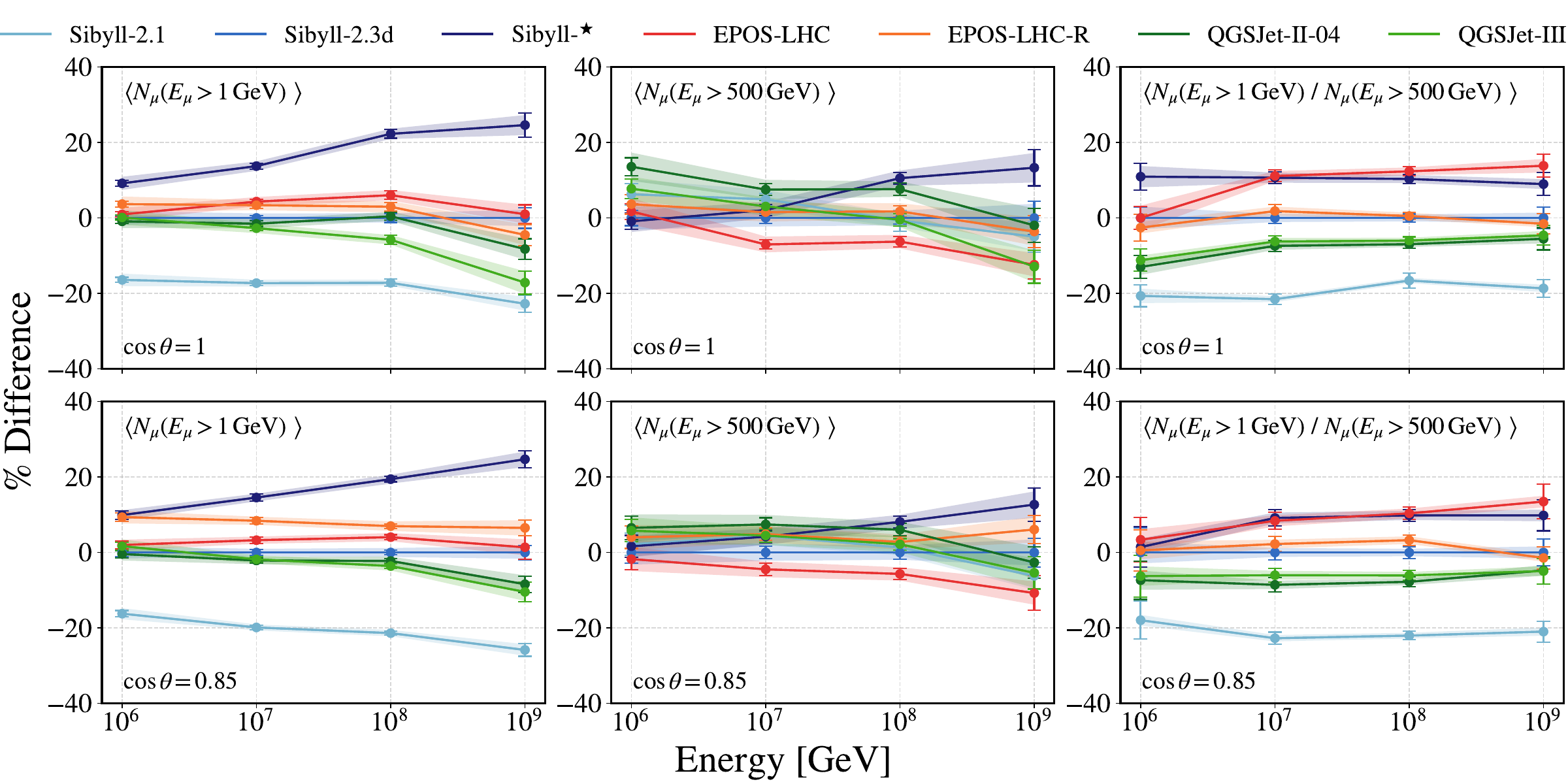}
\caption{Differences of hadronic interaction models with respect to Sibyll\,2.3d in the average muon number at the ground with $E_\mu>1$\,GeV (left) and $E_\mu>500$\,GeV (center) and the ratio of low-energy to high-energy muons (right) weighted according to the GSF flux model. Shaded bands indicate the uncertainties of the GSF flux model, while error bars represent statistical uncertainties. The distributions are shown for $\cos(\theta)=1$ (top) and $\cos(\theta)=0.85$ (bottom).}
\label{fig:diff_ratio_percent}
\end{figure}

\Cref{fig:contour_GSF} shows the 68.3\% contours (derived from kernel density estimation) of the number of muons, $N_\mu$, at the ground per simulated air shower with energies $E_\mu>1$\,GeV as a function of the number of muons with energies $E_\mu>500$\,GeV, obtained from the different hadronic models and using GSF mass fractions. 
The average model differences for low-energy and high-energy muons are depicted in \cref{fig:diff_ratio_percent} (left and center). The shaded bands indicate the uncertainties resulting from the composition in the GSF model, while the error-bars show the statistical uncertainties. For low-energy muons they range from around 30\% at low EAS energies, up to 50\% for EAS at $10^9$\,GeV. In particular, the muon number in Sibyll\,2.1 and Sibyll$^\bigstar$ differs significantly, while the other models only show deviations below 10-20\%. For the number of high-energy muons the models agree within 20\% over the entire energy range. Evaluating the muon number at various lateral distances instead of the total muon number on the ground yields similar results, both in terms of differences between hadronic models and in the dependence on the primary particle. The ratio of the number of high- to low-energy muons carries information about the steepness of the energy spectra. The model differences of this ratio are also shown in \cref{fig:diff_ratio_percent} (right). They are approximately constant for all air shower energies, ranging up to around 40\%. While the GeV muon content can distinguish the most extreme models, such as Sibyll\,2.1 and Sibyll$^\bigstar$, the ratio of low- to high-energy muons is necessary to effectively separate between EPOS-LHC, Sibyll\,2.3d, and QGSJET-II-04. QGSJet-III can potentially be constrained by muons in $10^9$\,GeV showers. EPOS-LHC-R can most clearly be distinguished by the number of GeV muons, the predicted muon ratios are similarly to those of Sibyll\,2.3d. %In general, distinguishing between models requires the detector resolution to be comparable to the differences predicted by the models. 
To effectively constrain all models, both the number of low energy muons and their ratio to high-energy muons must be simultaneously measured within the same air shower.

\section{Muon Production Depth} 
\label{sec:MPD}

The Pierre Auger Observatory has reported a measurement of the muon production depth in EASs~\cite{PierreAuger:2014zay}, employing data from surface detectors positioned far from the shower core ($>1700$\,m). While IceCube did not report any measurement of the muon production depth, new methods based on gradient boosted decision trees~\cite{Kravka:2023tyf} allow to use signals closer to the shower axis. This approach may enable a measurement of the muon production depth using IceTop data. In coincidence with a measurement of low- and high-energy muons, such measurements can provide strong constraints on hadronic interaction models.

\begin{figure}[!t]
\vspace{-1.em}
\mbox{\hspace{-1.em}\includegraphics[width=0.52\textwidth]{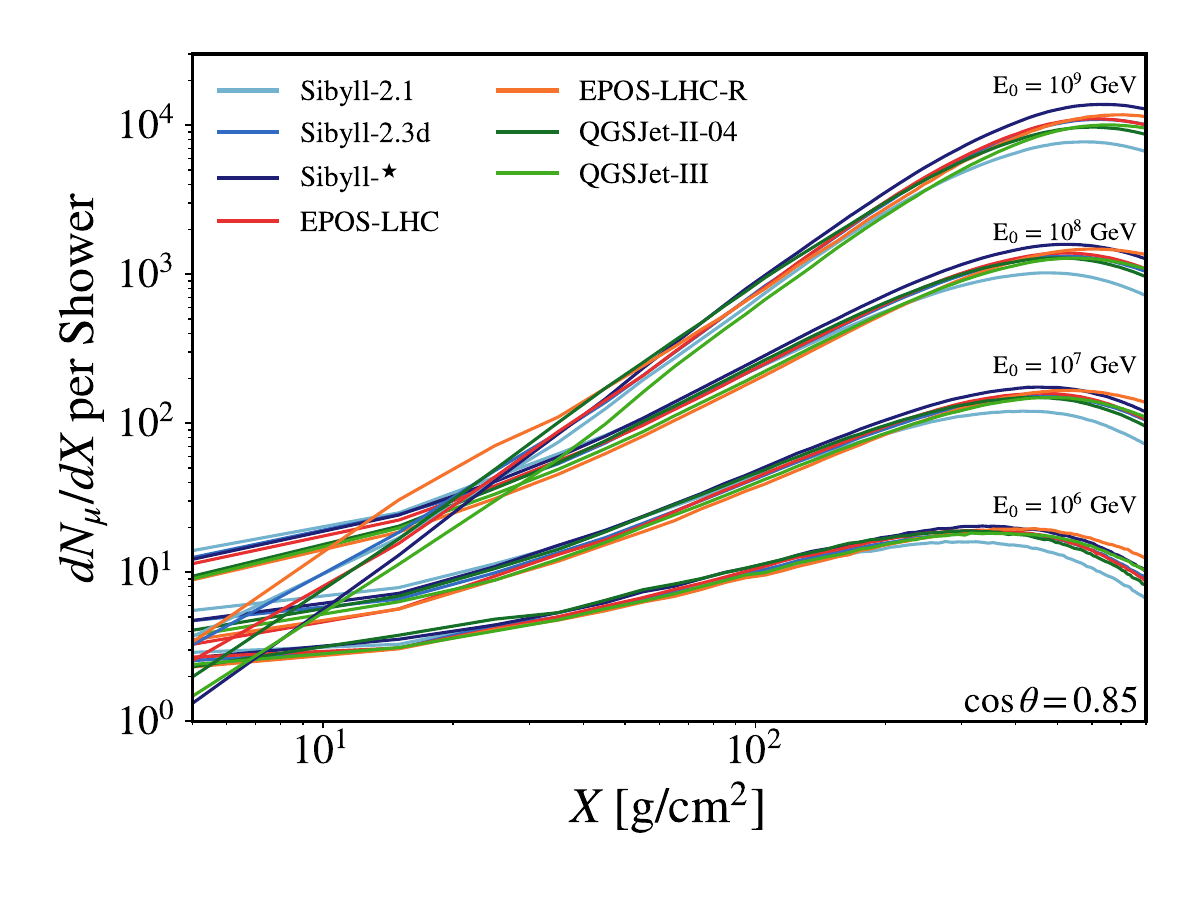}}
\includegraphics[width=0.52\textwidth]{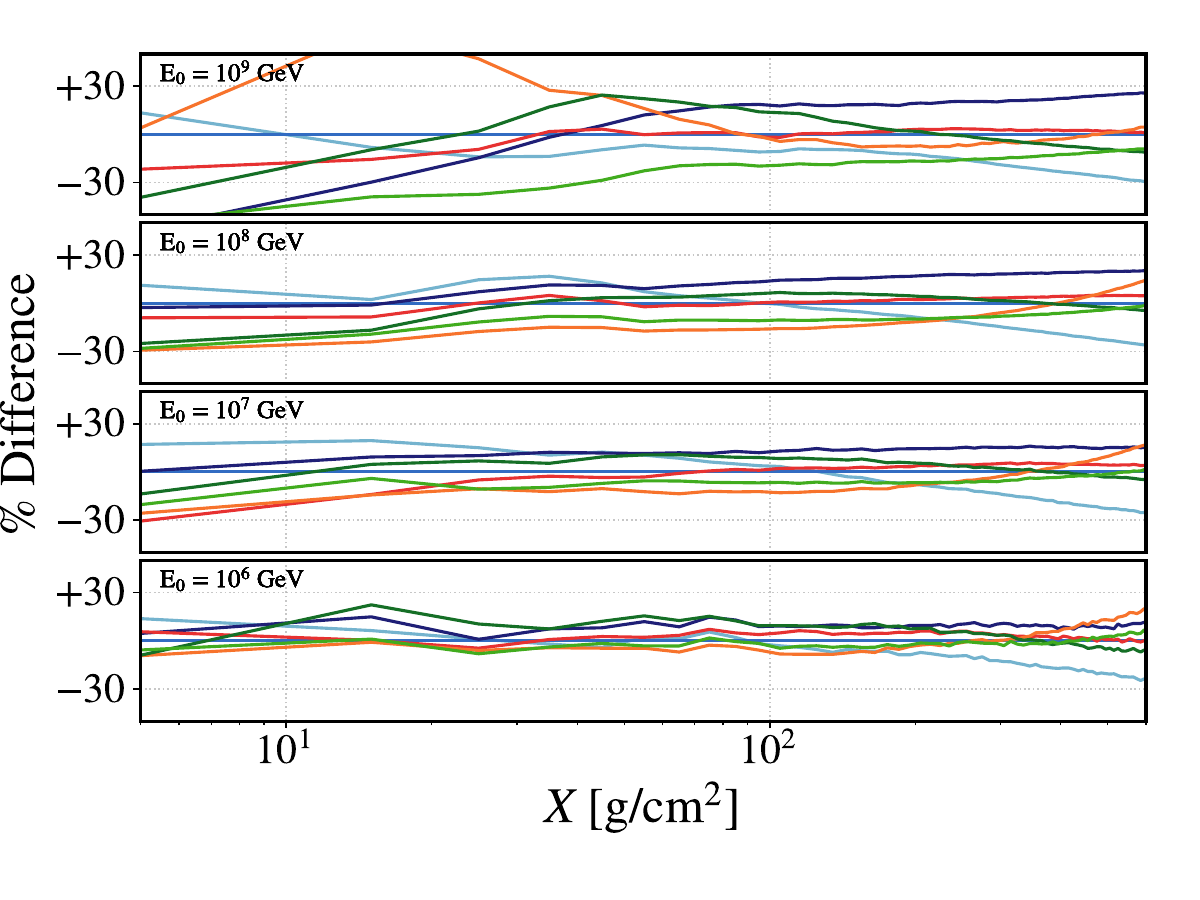}
\vspace{-2.em}
\caption{Average apparent muon production depth distributions for showers where $X_\mu^\mathrm{max}$ is above ground (left) obtained from CORSIKA simulations for hydrogen, helium, oxygen, iron primaries at $10^6$, $10^7$, $10^8$, $10^9$\,GeV and $\cos(\theta)=0.85$, using various hadronic interaction models. The primary mass fractions from the GSF model are used. Also shown are the model differences with respect to Sibyll\,2.3d for each primary energy separately (right).}
\label{fig:MPD}
\end{figure}

\begin{figure}[!b]
\centering
\vspace{-1em}
\includegraphics[width=\textwidth]{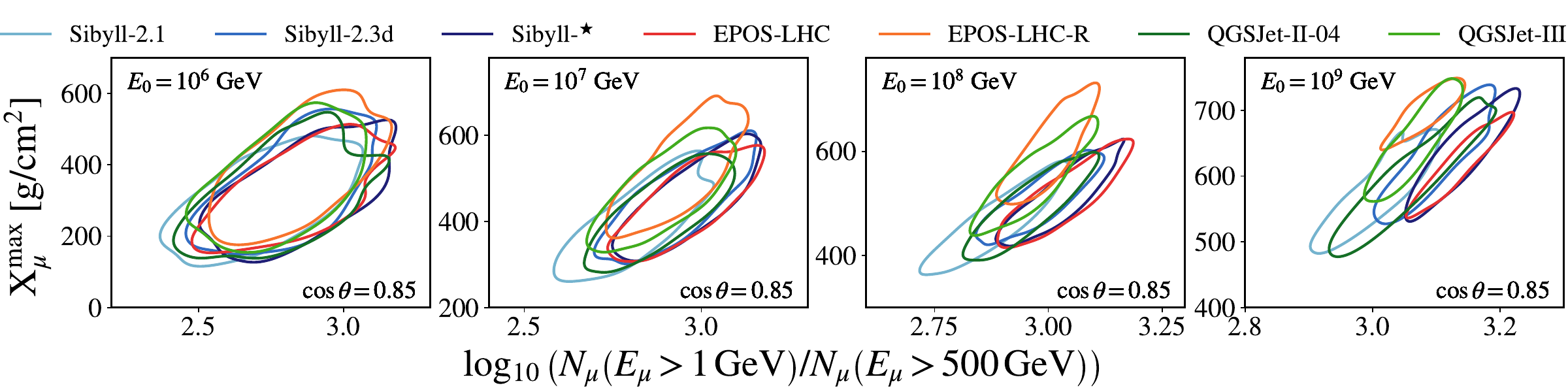}
\caption{68.3\% contours (derived from kernel density estimation) of the maximum apparent muon production depth, $X_\mu^\mathrm{max}$ for showers where $X_\mu^\mathrm{max}$ is above ground, as a function of the ratio of low-energy to high-energy muons per simulated air shower, obtained from the different hadronic models. The showers are weighted according to the GSF flux model.}
\label{fig:mu_prod_depth_scatter_GeV_to_500GeV}
\end{figure}
In the following, the apparent distributions of the production depth is used, considering only the production depth of muons that reach the ground and have an energy > 1\,GeV. Muons that decay between production and the observation level are not included. Furthermore, only showers in which the maximum of the apparent muon production depth occurs above ground are selected. \cref{fig:MPD} (left) shows the apparent muon production depths for air showers from zenith angle directions of $\cos(\theta)=0.85$, and primary energies of $10^6$, $10^7$, $10^8$, and $10^9$\,GeV. These distributions are obtained from air shower simulations using hydrogen, helium, oxygen, and iron primaries, and their mass composition is included based on to the GSF flux model. The corresponding differences between the simulated profiles based on the different models with respect to Sibyll\,2.3d are shown in \cref{fig:MPD} (right). At higher energies, the maximum of the apparent muon production depth is typically below the ground, at around $690,\mathrm{g/cm}^2$. Protons interact deeper in the atmosphere than iron nuclei, due to their smaller cross-section, increasing the likelihood that $X_\mu^\mathrm{max}$ cannot be reconstructed, as it lies below ground, which introduces a different composition bias at higher energies. Considering more inclined showers alleviates this bias, as more atmosphere has to be traversed and $X_\mu^\mathrm{max}$ remains above ground. 

\begin{figure}[tb]
%\vspace{-1em}
\centering
\includegraphics[width=\textwidth]{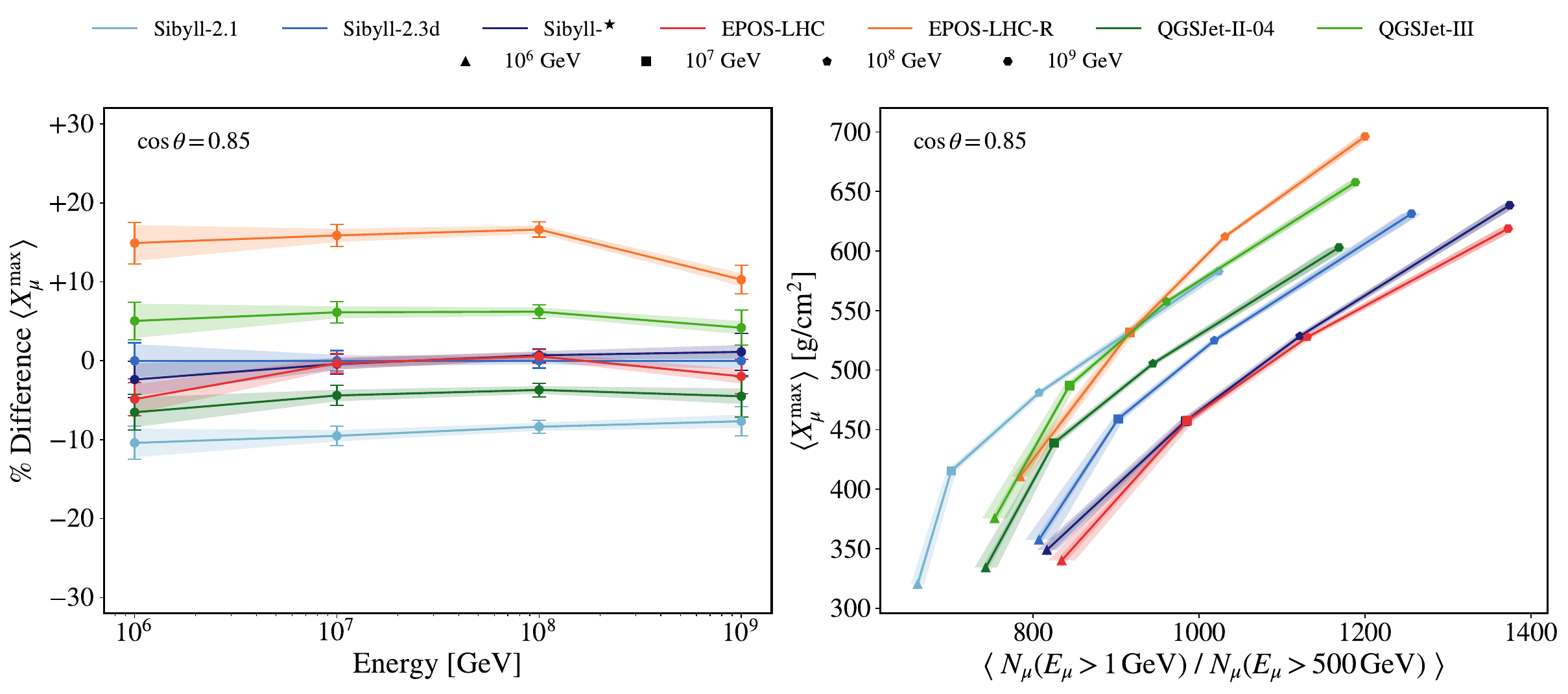}
\caption{Left: Differences of hadronic interaction models with respect to Sibyll\,2.3d in the average maximum muon production depth, $\langle X_\mu^\mathrm{max}\rangle$ for showers where $X_\mu^\mathrm{max}$ is above ground. The shaded bands represent the uncertainties derived from the GSF flux model, while the error-bars show the statistical uncertainty. Right: Average apparent maximum production depth, $\langle X_\mu^\mathrm{max}\rangle$, as a function of the average ratio of high- to low-energy for showers where $X_\mu^\mathrm{max}$ is above ground. The distributions are shown for various hadronic interaction models and EAS energies. The shaded bands represent the uncertainties derived from the GSF flux model. }
\label{fig:mu_prod_depth_GSF}
\vspace{-1em}
\end{figure}

\Cref{fig:mu_prod_depth_scatter_GeV_to_500GeV} shows the 68.3\% contours (derived from kernel density estimation) of the maximum of the apparent muon production depth, $X_\mu  ^\mathrm{max}$, per simulated air shower as a function of the ratio of the number of high- and low-energy muons, obtained from the different hadronic models and using GSF mass fractions. The model differences in the average maximum muon production depth, $\langle X_\mu^\mathrm{max}\rangle$, reach up to around 30\%, as depicted in \cref{fig:mu_prod_depth_GSF} (left).

\newpage
The average apparent maximum production depth is also shown in \cref{fig:mu_prod_depth_GSF} on the right, as a function of the ratio of high- to low-energy muons, for various hadronic interaction models and different EAS energies. A measurement of the apparent muon production depth (in particular in more horizontal showers) in coincidence with the high- and low-energy muon numbers provides additional constraints on hadronic interaction models. Specifically, it will allow to distinguish between the three Sibyll models, as well as between the two EPOS models. Thereby, the measurement of multiple muon observables in the same air shower will provide stringent constraints on hadronic interaction models.

\section{Conclusions}

We have presented a comprehensive simulation study of the muon content in EASs. It was shown that a simultaneous measurement of low-energy and high-energy muons in the same air shower provides unique opportunities to test hadronic interaction models. In particular, their ratio probes the energy spectrum of muons and gives insights into the energy sharing in EASs, which enables distinctive tests of hadronic model predictions, including most recent ones. Predictions of this ratio differ by up to 50\% over the entire EAS energy range considered, exceeding the expected precision for such measurements with existing experiments. Dedicated analyses can be performed with IceCube data in coincidence with data from its surface array detector, IceTop, for example, or with the proposed IceCube-Gen2 experiment~\cite{IceCube-Gen2:2020qha}. When combined with estimates of the maximum muon production depth, these measurements may be able to provide even more stringent constraints on a large variety of EAS models in the future.

%\newpage
\section*{Acknowledgements}

The authors acknowledge the support and resources from the Center for High Performance Computing at the University of Utah. This work was supported by a Postdoc Travel Assistance Award awarded to L.\,P. from the Postdoc Affairs Office at the University of Utah. F.\,R. acknowledges funding from the German Science Foundation (DFG), via the Collaborative Reasearch Center SFB1491 ``Cosmic Interacting Matters - From Source to Signal''.

\bibliographystyle{ICRC}
\bibliography{references}

\providecommand{\href}[2]{#2}\begingroup\raggedright\begin{thebibliography}{10}

\bibitem{Coleman:2022abf}
A.~Coleman {\em et~al.}, \href{http://dx.doi.org/10.1016/j.astropartphys.2023.102819}{{\em Astropart. Phys.} {\bfseries 149} (2023) 102819}.

\bibitem{PierreAuger:2014ucz}
{\bfseries Pierre Auger} Collaboration, A.~Aab {\em et~al.}, \href{http://dx.doi.org/10.1103/PhysRevD.91.032003}{{\em Phys. Rev. D} {\bfseries 91} no.~3, (2015) 032003}. [Erratum: Phys.Rev.D 91, 059901 (2015)].

\bibitem{PierreAuger:2016nfk}
{\bfseries Pierre Auger} Collaboration, A.~Aab {\em et~al.}, \href{http://dx.doi.org/10.1103/PhysRevLett.117.192001}{{\em Phys. Rev. Lett.} {\bfseries 117} no.~19, (2016) 192001}.

\bibitem{PierreAuger:2024neu}
{\bfseries Pierre Auger} Collaboration, A.~Abdul~Halim {\em et~al.}, \href{http://dx.doi.org/10.1103/PhysRevD.109.102001}{{\em Phys. Rev. D} {\bfseries 109} no.~10, (2024) 102001}.

\bibitem{Dembinski:2019uta}
{\bfseries EAS-MSU, IceCube, KASCADE Grande, NEVOD-DECOR, Pierre Auger, SUGAR, Telescope Array, Yakutsk EAS Array} Collaboration, H.~P. Dembinski {\em et~al.}, \href{http://dx.doi.org/10.1051/epjconf/201921002004}{{\em EPJ Web Conf.} {\bfseries 210} (2019) 02004}.

\bibitem{Soldin:2021wyv}
{\bfseries EAS-MSU, IceCube, KASCADE-Grande, NEVOD-DECOR, Pierre Auger, SUGAR, Telescope Array, Yakutsk EAS Array} Collaboration, D.~Soldin, \href{http://dx.doi.org/10.22323/1.395.0349}{{\em PoS} {\bfseries ICRC2021} (2021) 349}.

\bibitem{Cazon:2020zhx}
{\bfseries EAS-MSU, IceCube, KASCADE Grande, NEVOD-DECOR, Pierre Auger, SUGAR, Telescope Array, Yakutsk EAS Array} Collaboration, L.~Cazon, \href{http://dx.doi.org/10.22323/1.358.0214}{{\em PoS} {\bfseries ICRC2019} (2020) 214}.

\bibitem{Albrecht:2021cxw}
J.~Albrecht {\em et~al.}, \href{http://dx.doi.org/10.1007/s10509-022-04054-5}{{\em Astrophys. Space Sci.} {\bfseries 367} no.~3, (2022) 27}.

\bibitem{ArteagaVelazquez:2023fda}
J.~C. Arteaga~Velazquez, \href{http://dx.doi.org/10.22323/1.444.0466}{{\em PoS} {\bfseries ICRC2023} (2023) 466}.

\bibitem{Aartsen:2016nxy}
{\bfseries IceCube} Collaboration, M.~G. Aartsen {\em et~al.}, \href{http://dx.doi.org/10.1088/1748-0221/12/03/P03012}{{\em JINST} {\bfseries 12} no.~03, (2017) P03012}.

\bibitem{Verpoest:2023qmq}
{\bfseries IceCube} Collaboration, S.~Verpoest, \href{http://dx.doi.org/10.22323/1.444.0207}{{\em PoS} {\bfseries ICRC2023} (2023) 207}.

\bibitem{IceCubeCollaboration:2022tla}
{\bfseries IceCube} Collaboration, R.~Abbasi {\em et~al.}, \href{http://dx.doi.org/10.1103/PhysRevD.106.032010}{{\em Phys. Rev. D} {\bfseries 106} no.~3, (2022) 032010}.

\bibitem{IceCube:2023suf}
{\bfseries IceCube} Collaboration, M.~Weyrauch and D.~Soldin, \href{http://dx.doi.org/10.22323/1.444.0357}{{\em PoS} {\bfseries ICRC2023} (2023) 357}.

\bibitem{Mark:ICRC2025}
{\bfseries IceCube} Collaboration, M.~Weyrauch, {\em PoS} {\bfseries ICRC2025} (2025) 437.

\bibitem{Lincoln:ICRC2025}
{\bfseries IceCube} Collaboration, L.~Draper, F.~Varsi, and D.~Soldin, {\em PoS} {\bfseries ICRC2025} (2025) 424.

\bibitem{LillyICRC25}
{\bfseries IceCube} Collaboration, L.~Pyras, {\em PoS} {\bfseries ICRC2025} (2025) 367.

\bibitem{Dembinski:2017zsh}
H.~P. Dembinski, R.~Engel, A.~Fedynitch, T.~K. Gaisser, F.~Riehn, and T.~Stanev, \href{http://dx.doi.org/10.22323/1.301.0533}{{\em PoS} {\bfseries ICRC2017} (2018) 533}.

\bibitem{Heck:1998vt}
D.~Heck, J.~Knapp, J.~N. Capdevielle, G.~Schatz, and T.~Thouw, \href{http://dx.doi.org/10.5445/IR/270043064}{{\em FZKA Report 6019} (1998) }.

\bibitem{Ahn:2009wx}
E.-J. Ahn, R.~Engel, T.~K. Gaisser, P.~Lipari, and T.~Stanev, \href{http://dx.doi.org/10.1103/PhysRevD.80.094003}{{\em Phys. Rev. D} {\bfseries 80} (2009) 094003}.

\bibitem{Riehn:2019jet}
F.~Riehn, R.~Engel, A.~Fedynitch, T.~K. Gaisser, and T.~Stanev, \href{http://dx.doi.org/10.1103/PhysRevD.102.063002}{{\em Phys. Rev. D} {\bfseries 102} no.~6, (2020) 063002}.

\bibitem{Pierog:2013ria}
T.~Pierog, I.~Karpenko, J.~M. Katzy, E.~Yatsenko, and K.~Werner, \href{http://dx.doi.org/10.1103/PhysRevC.92.034906}{{\em Phys. Rev. C} {\bfseries 92} no.~3, (2015) 034906}.

\bibitem{Pierog:2023ahq}
T.~Pierog and K.~Werner, \href{http://dx.doi.org/10.22323/1.444.0230}{{\em PoS} {\bfseries ICRC2023} (2023) 230}.

\bibitem{Ostapchenko:2013pia}
S.~Ostapchenko, \href{http://dx.doi.org/http://dx.doi.org/10.1051/epjconf/20125202001}{{\em EPJ Web Conf.} {\bfseries 52} (2013) 02001}.

\bibitem{Ostapchenko:2005nj}
S.~Ostapchenko, \href{http://dx.doi.org/10.1103/PhysRevD.74.014026}{{\em Phys. Rev. D} {\bfseries 74} no.~1, (2006) 014026}.

\bibitem{Ostapchenko:2024myl}
S.~Ostapchenko, \href{http://dx.doi.org/10.1103/PhysRevD.109.094019}{{\em Phys. Rev. D} {\bfseries 109} no.~9, (2024) 094019}.

\bibitem{Riehn:2024prp}
F.~Riehn, A.~Fedynitch, and R.~Engel, \href{http://dx.doi.org/10.1016/j.astropartphys.2024.102964}{{\em Astropart. Phys.} {\bfseries 160} (2024) 102964}.

\bibitem{PierreAuger:2014zay}
{\bfseries Pierre Auger} Collaboration, A.~Aab {\em et~al.}, \href{http://dx.doi.org/10.1103/PhysRevD.90.012012}{{\em Phys. Rev. D} {\bfseries 90} no.~1, (2014) 012012}. [Addendum: Phys.Rev.D 90, 039904 (2014), Erratum: Phys.Rev.D 92, 019903 (2015)].

\bibitem{Kravka:2023tyf}
A.~Kravka, E.~Santos, M.~Stadelmaier, and A.~Yushkov, \href{http://dx.doi.org/10.22323/1.444.0282}{{\em PoS} {\bfseries ICRC2023} (2023) 282}.

\bibitem{IceCube-Gen2:2020qha}
{\bfseries IceCube-Gen2} Collaboration, M.~G. Aartsen {\em et~al.}, \href{http://dx.doi.org/10.1088/1361-6471/abbd48}{{\em J. Phys. G} {\bfseries 48} no.~6, (2021) 060501}.

\end{thebibliography}\endgroup

\end{document}